\documentclass[showpacs,twocolumn,floatfix]{revtex4}
\usepackage{amsmath}
\usepackage{amsfonts}
\usepackage{amssymb}
\usepackage{graphicx}
\usepackage{color}
\begin{document}
\title{Energy transport in a one-dimensional granular gas}
\author{\'Italo'Ivo Lima Dias Pinto$^1$}
\author{Alexandre Rosas$^1$}
\author{Katja Lindenberg$^{2}$}
\affiliation{${}^1$ Departamento de F\'{\i}sica, CCEN, Universidade Federal da Para\'{\i}ba, Caixa Postal 5008, 58059-900, Jo\~ao Pessoa, Brazil\\
${}^2$Department of Chemistry and Biochemistry,
and Institute for Nonlinear Science, University of
California San Diego, La Jolla, CA 92093-0340}

\begin{abstract}
We study heat conduction in one-dimensional granular gases. In particular, we consider two
mechanisms of viscous dissipation during inter-grain collisions. In one, the dissipative force
is proportional to the grain's velocity and dissipates not only energy but also momentum. In
the other, the dissipative force is proportional to the relative velocity of the grains and
therefore conserves momentum even while dissipating energy. 
This allows us to explore the role of momentum conservation in the heat conduction properties of
this one-dimensional nonlinear system. We find normal thermal conduction whether or not momentum is
conserved.
\end{abstract}
\pacs{45.70.-n,05.20.Dd,05.70.Ln,83.10.Rs}

\maketitle

\section{Introduction}
\label{introduction}
Energy transport in low-dimensional systems can be pathological in the sense that Fourier's law for heat
conduction might break down~\cite{narayan02, lepri03,dhar08,lepri03b}. In three-dimensional solids,
the energy (heat) transported is governed by Fourier's law, which says that the heat flux $J$ 
is proportional to the gradient of the temperature,
\begin{equation}
J = -\kappa \nabla T,
\end{equation}
where $\kappa$ is the thermal conductivity.  The thermal conductivity in Fourier's law is independent
of system size and of time. This equation assumes
that a local equilibrium is established at each time, so that one can define the local energy flux $J(x,t)$
and temperature $T(x,t)$, when the temperature gradient lies along the $x$
direction~\cite{lepri03b}.
In one-dimensional systems one might expect a similar equation, with the gradient replaced by the
derivative of $T(x,t)$ with respect to $x$, and $J$ simply being the energy flux per unit time.
However, it has been observed that in many one-dimensional models the thermal conductivity varies
with the system size $N$ as  $\kappa \sim N^{\alpha}$, with $\alpha$ lying between 0.32 and
0.44~\cite{lepri03b,mai06,mai07}.
Self-consistent mode coupling analysis for one-dimensional nonlinear chains~\cite{lepri98} predicts a
universal value of $\alpha=2/5$, while renormalization group analysis based on hydrodynamical models
supports a universality class with $\alpha=1/3$. Recently, Mai et al.~\cite{mai06,mai07} revisited
the problem and obtained the same $\alpha=1/3$ behavior for one-dimensional nonlinear chains as
obtained for the fluid-like systems. They argue that the dynamic equations are those of a fluid at
{\em all} length scales even if the static order of the chain extends to very large $N$.
They explain the disagreement with most numerical results as a consequence of the numerical
difficulties in reaching asymptotic behavior, which requires extremely large $N$, but their
conclusions have in turn been disputed~\cite{mai07}.
Whether there is a single universality class or there are two, there is a connection between
heat conduction and diffusion in one-dimensional systems~\cite{li03}, that is, normal diffusion leads
to normal heat conductivity ($\kappa \sim N^0$), whereas anomalous transport is associated with
$\kappa \sim N^\alpha$, with $\alpha>0$ ($<0$) for super(sub)diffusion.  Thus,
superdiffusion is associated with a divergent thermal conductivity. 

Another key concept in the understanding of the thermal conductivity in $1d$ systems is momentum
conservation~\cite{lepri03b}.  In fact, momentum conservation usually implies the divergence of the
thermal conductivity in $1d$~\cite{narayan02,prosen00},
and yet it is not necessary for the occurrence of anomalous diffusion~\cite{prosen05}. An exception to
the momentum conservation rule is a chain of particles interacting via the nearest-neighbor potential
$V(q_i - q_{i-1}) = 1 - \cos (q_i - q_{i-1})$, the so-called rotator model~\cite{giardina00}.
This momentum-conserving model exhibits normal transport behavior to very high numerical accuracy,
and is said to occur because the rotator model ``cannot support a nonvanishing pressure, and thus
infinite-wavelength phonons cannot carry energy''~\cite{prosen05}. In~\cite{lepri03b} this behavior
is ascribed to the periodicity of the potential and the associated independent jumps from valley to
valley.

In this contribution we discuss heat conduction in one-dimensional granular gases.   
Granular gases are dissipative systems, the associated energy loss usually being modeled by introducing a
coefficient of restitution as a parameter in the description of granular collisions.  Instead, we
follow a more dynamical approach and introduce energy dissipation explicitly via viscous terms in the
equations of motion for the granules. There are a number of different sources and descriptions of
viscous effects in the literature~\cite{duvall69,brunhuber06,arevalo02,brilliantov,sen,falcon05,our1}.
The different ways in which they appear in the dynamical descriptions suits our particular focus of
interest, which is the role of momentum conservation in the heat transport process.  In particular,
one way to introduce viscosity conserves momentum~\cite{ourprl,ourpre,herbold1,herbold2},
while the other does not~\cite{our1,ourgas,our2}.  Interestingly, we establish that thermal conduction
exhibits normal behavior in both cases, that is, that there is no divergence as the system size
increases.

In Sec.~\ref{models} we describe the two dissipative models, and in Sec.~\ref{simulations} we
present simulation results to characterize the transport of heat through the granular gas in both
cases.  We conclude with a short summary in Sec.\ref{conclusions}.

\section{Dissipative Models}
\label{models}

We consider $N$ identical granules constrained to move on a line between two walls at different temperatures.
The granules move freely except during collisions either with the walls or with one another. 
The system is a one-dimensional ``granular gas" because the distance between the walls is much
greater than the space occupied by the granules.
The walls act as heat baths, that is, whenever a granule collides with a wall at temperature $T$, its
energy is absorbed by the wall and it acquires a new velocity away from the wall
according to the probability distribution~\cite{tehver98} 
\begin{equation}
P(v) = \frac{|v|}{T} \exp \left( -v^ 2/2k_BT \right).
\label{distribution}
\end{equation}
Interparticle collisions are governed by the power-law potential
\begin{equation}
\begin{array}{l l l}
V(\delta_{k,k+1})&=\frac{\displaystyle a}{\displaystyle n}|\delta|^{n}_{k,k+1}, \qquad &\delta\leq 0,\\ \\
V(\delta_{k,k+1})&= 0, \qquad &\delta >0.
\end{array}
\label{eq:hertz}
\end{equation}
Here
$\delta_{k,k+1} \equiv y_{k+1} - y_{k}$, $a$ is a prefactor determined by Young's modulus and Poisson's ratio, 
and the principal radius of curvature $R$ of the surfaces at the point of
contact~\cite{landau,hertz};
and $y_k$ is the displacement of granule $k$ from its position at the
beginning of the collision.
The exponent $n$ is $5/2$ for spheres (Hertz potential), it is $2$ for cylinders, and
in general depends on geometry. 
We stress that the one-sided (only repulsion) granular potential even with
$n=2$ is entirely different from a two-sided (repulsion and attraction)
harmonic potential.
The one-sidedness of the potential leads to
analytic complexities even in the dissipationless
case~\cite{our2,nesterenko,hinch}, and even greater complexities in the
presence of dissipation~\cite{our1,ourprl,ourpre,our2}.

In this paper we 
explore the \emph{low density} limit, which leads to enormous
analytical and computational simplifications.
The low density feature of these approximations is implemented via the
assumption that the collisions are always \emph{binary}~\cite{ourbinary}, that is, that only
two granules at a time are members of any collision event, and that at any
moment of time there is at most one collision.  

Our approach starts with the equations of motion of the particles labeled by index $k$, and so
we write $y_k$ as a function of time $\tau$. It is convenient to deal
with scaled position and time variables $x_k$ and $t$, related to 
the unscaled variables $y_k$ and $\tau$ as follows,
\begin{equation}
y_k = \left( \frac{m v_0^2}{a} \right) ^{1/n} x_k, \qquad
\tau = \frac{1}{v_0} \left( \frac{m v_0^2}{a} \right) ^{1/n} t. 
\end{equation}
Here $m$ is the mass of the granules, and
the velocity $v_0$ is an arbitrary choice in terms of which other velocities are expressed. We also
introduce a scaled friction coefficient 
\begin{equation}
\gamma =\frac{\tilde{\gamma}}{m v_0} \left( \frac{m v_0^2}{a} \right)
^{1/n},
\end{equation}
where $\tilde{\gamma}$ is the friction coefficient for the unscaled variables.
In the low density limit we only need to consider 
the equations of motion for two colliding granules $k=1,2$.  Furthermore, in this paper we only
consider cylindrical grains, which leads to considerable simplification while still capturing
the important general features of the system that we seek to highlight. We stress that the
one-sided granular potential (i.e., one with only repulsive interactions) even with $n=2$ is
entirely different from a two-sided harmonic potential. 

Consider a viscous force that is proportional to the relative velocity of the colliding granules.
Such a force has been considered not only theoretically~\cite{duvall69,ourprl,ourpre} but it has
also been observed experimentally~\cite{herbold1,herbold2}.  While one might be tempted to think of
this force as arising because one grain rubs against the other, the actual mechanism is more
complicated and involves the medium that surrounds the granules~\cite{herbold1,herbold2}. The exact
mechanism is still a matter of conjecture. In any case, the 
appropriate equations of motion in this case are 
\begin{eqnarray}
  \ddot{x}_1 &=& \left [-\gamma (\dot{x}_1 - \dot{x}_{2}) - (x_1 - x_{2})\right] \theta (x_1 -
x_{2}), \nonumber \\
  \ddot{x}_2 &=& \left [\gamma (\dot{x}_1 - \dot{x}_{2}) + (x_1 - x_{2})\right] \theta (x_1 -
x_{2}),
\label{eq:motion_conserved}
\end{eqnarray}
where a dot denotes a derivative with respect to $t$. 
The Heaviside function is defined as $\theta(x)=1$ for $x>0$, $\theta(x) =0$ for $x<0$, and
$\theta(0)=1/2$.  It ensures that the two particles interact only when in contact, that is, only when
the particles are loaded.  The post-collisional velocities (called $u$ below) can be written
in terms of the velocities of the two granules at the beginning of the collision (called $v$) as
\begin{eqnarray}
	u_1 &=& \frac{1}{2} \left[ 1 - e^{-\gamma t_0} \right] v_1 + \frac{1}{2} \left[ 1 +
e^{-\gamma t_0} \right] v_2, \nonumber \\
	u_2 &=& \frac{1}{2} \left[ 1 + e^{-\gamma t_0} \right] v_1 + \frac{1}{2} \left[ 1 -
e^{-\gamma t_0} \right] v_2, 
	\label{eq:post-vel-cons}
\end{eqnarray}
where $t_0 = \pi/(\sqrt{2-\gamma^2})$ is the duration of the collision. 
Since $u_1+u_2= v_1+v_2$, the momentum of the center of mass is
conserved. 

A model in which the center of mass velocity is not conserved is one governed by the equations
of motion~\cite{our1,our2}
\begin{eqnarray}
\ddot{x}_1 &=& \left[-\gamma \dot{x}_1 - (x_1 - x_{2})\right] \theta (x_1 -
x_{1}),\nonumber \\
\ddot{x}_2 &=& \left[-\gamma \dot{x}_2 + (x_1 - x_{2})\right] \theta (x_1 -
x_{1}).
\label{eq:motion_rescaled}
\end{eqnarray}
Here one might think of the viscosity arising from an interaction with the medium.  However, such an
interaction would produce a damping term not only during a collision but also while the granules are
moving independently between collisions.  In the low density limit the granules would hardly collide
before stopping entirely unless the viscosity is extremely low, and one can then not talk of heat
transport along the one-dimensional system.  Thus, this model, while perhaps not realistic in any sense,
is a ``toy model" in which momentum is not conserved while still capable of supporting energy transport 
and therefore relevant to our question.  Indeed, we can again calculate the post-collision velocities,
\begin{eqnarray}
	u_1 &=& \frac{1}{2} \left[ e^{-\gamma t_0} - e^{-\gamma t_0/2} \right] v_1 + \frac{1}{2} \left[ e^{-\gamma t_0} + e^{-\gamma t_0/2} \right] v_2, \nonumber \\
	u_2 &=& \frac{1}{2} \left[ e^{-\gamma t_0} + e^{-\gamma t_0/2} \right] v_1 + \frac{1}{2} \left[ e^{-\gamma t_0} - e^{-\gamma t_0/2} \right] v_2,
	\label{eq:post-vel-diss}\nonumber\\
\end{eqnarray}
where $t_0 = 2\pi/(\sqrt{8-\gamma^2})$ is again the duration of the collision. It is easy to verify that
in this case the momentum of the center of mass decays from $v_1+v_2$ before a collision to 
\begin{equation}
u_1 + u_2 = \left[ v_1 + v_2 \right] e^{-\gamma t_0}, 
\end{equation}
after the collision, so that the momentum is not conserved.

Caution must be exercised in the choice of the damping coefficient $\gamma$.  If it is too large,
energy acquired from either wall is simply dissipated before it crosses the system, in which case the
problem changes from one of energy transfer from one wall to the other through the granular gas
to that of two walls pumping energy into the granular ``sink."  To estimate the
limit on the damping coefficient we can imagine a sequence of collisions whereby a granule starting
from the hot wall with kinetic energy $E_0$ collides with the next granule, which in turn collides
with the next one, and so on, until the last granule, whose energy is $E_N$,  collides with the cold
wall.  The change in the kinetic energy due to each collision in the momentum conserving model is 
\begin{equation}
\Delta E \equiv \frac{1}{2}(u_1^2 +u_2^2) - \frac{1}{2} (v_1^2+v_2^2) = -\frac{1}{4} \left(1-e^{-2\gamma t_0}\right)
(v_1+v_2)^2,
\end{equation}
and in the momentum non-conserving model it is
\begin{equation}
\Delta E \equiv \frac{1}{2}(u_1^2 +u_2^2) - \frac{1}{2} (v_1^2+v_2^2) = -\frac{1}{4} \left(1-e^{-\gamma t_0}\right)
(v_1+v_2)^2.
\end{equation}
In both cases the kinetic energy before and after the collision are related by an expression of the
form $E_{after} = E_{before} (1-b\gamma t_0)$ where $b$ is a velocity-dependent dimensionless
quantity of order unity, and where we have assumed that $\gamma t_0 \ll 1$.  As an order of magnitude
estimate it is sufficient to write
$E_N \sim E_0 (1-\gamma t_0) ^N$.  The average energy of a hot granule is $k_BT_1/2$ and in order
for energy to flow across the system the energy of the granule that arrives at the cold wall must be
greater than $k_B T_2/2$. Consequently, we estimate that a requirement for energy transport is that
\begin{equation}
 \gamma \lesssim 1 - \left( \frac{T_2}{T_1} \right)^{1/N}.
\label{delimit}
\end{equation}

We now wish to explore the following specific questions.  Is there a finite thermal conductivity
for either or both of these models?  If so, which model has a higher/lower thermal conductivity?
How does the thermal conductivity depend on temperature? On viscosity? On system size?  
These are the questions we address in the next section.

\section{Numerical Simulations}
\label{simulations}

\begin{figure}[ht]
	\begin{center}
		\includegraphics[width=6cm, angle=-90]{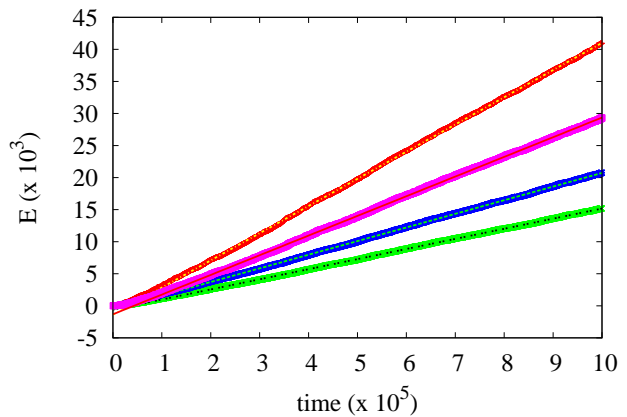}
	\end{center}
	\caption{(Color online) From top to bottom: energy injection by the hot wall (red), energy
absorption by the cold wall (magenta), energy injection by the cold wall (blue), and 
energy absorption by the hot wall (green),
for a momentum non-conserving system of $N=500$ particles between walls
maintained at $T_l = 6$ and $T_r=3$ and with $\gamma=0.0002$. The dashed lines are the linear
regressions of these curves (after the transient). They yield the rate of energy injection/absorption.
The initial temperature of the granular gas is $T=1$.}
	\label{fig:energy-cons}
\end{figure}

The low density limit allows us to use an event driven algorithm in our simulations.
As indicated earlier, when a granule collides with a wall at temperature $T$ its energy is absorbed
by the wall and it acquires a new energy as determined by the velocity distribution given in
Eq.~(\ref{distribution}).
The initial temperature of the granular gas is arbitrarily set to $T=1$, and we allow the
system to arrive at a steady state before beginning our ``measurements."
While ideally one would want the temperature gradient in the steady state to be strictly linear as
assumed in the linear response theory that leads to the Fourier law, the subtleties encountered in
$1d$ systems are well-known and observed pretty much no matter how the thermalization is
implemented~\cite{tehver98,mai07} and the resulting gradients are only strictly linear away from the
boundaries (nonlinear behavior typically sets in near the boundaries).
In any case, we can address the question of whether or not there are system size divergences in the
transport of heat.  That we do in fact achieve a steady state can be seen, for example, 
in Fig.~\ref{fig:energy-cons}, where we show the energy absorbed
and injected by each wall as a function of time for the dissipating momentum model.  After a 
short transient, the rate of energy injection and absorption become constant, as they should in
a steady state.  We ran tests to ascertain
that the initial temperature of the granular gas is not important for this equilibration, that is,
we find that the rates of energy injection and absorption are independent of the initial velocity
distribution of the granular gas. The momentum-conserving model
equilibrates equally well, also independently of initial condition.
Henceforth we set the initial temperature of the granular gas to unity. 

\begin{figure}[ht]
	\begin{center}
		\includegraphics[width=6cm, angle=-90]{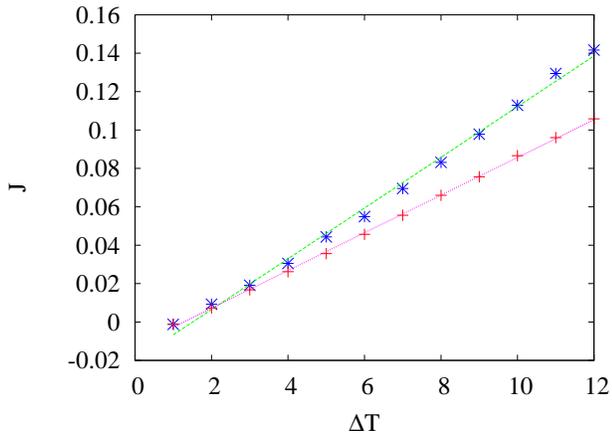}
	\end{center}
	\caption{(Color online) Rates of energy transmission as a function of the temperature gradient. In
this simulation the viscosity coefficient was set to $\gamma=0.0002$ and the system is composed of 500 granules.
The (red) plus signs correspond to the momentum conserving system, while the (blue) stars correspond to the momentum
dissipating system. In this figure, the cold wall temperature was set to 3.}
	\label{fig:fourier}
\end{figure}

\begin{figure}[ht]
	\begin{center}
		\includegraphics[width=6cm, angle=-90]{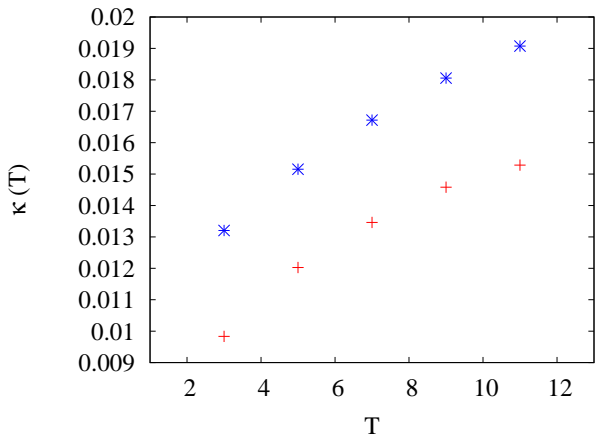}
	\end{center}
	\caption{(Color online) Thermal conductivity as a function of the temperature. In this simulation the viscosity
coefficient was set to $\gamma=0.0002$ and the system is composed of 500 granules. The (red) plus
signs correspond to the momentum conserving system, the (blue) stars to the momentum dissipating
system.}
	\label{fig:kappaT}
\end{figure}

\begin{figure}[ht]
	\begin{center}
		\includegraphics[width=6cm, angle=-90]{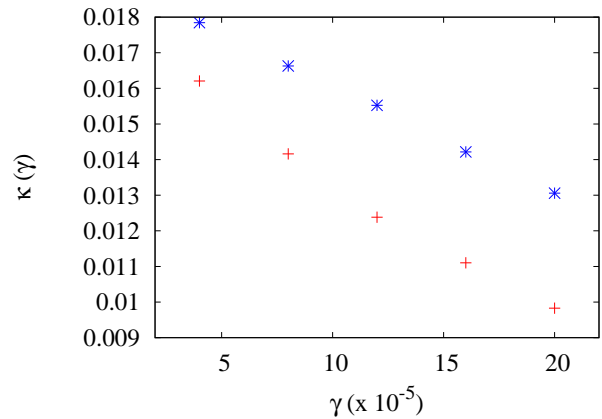}
	\end{center}
	\caption{(Color online) Thermal conductivity as a function of the viscosity coefficient. In this simulation
the temperature of the cold wall is set to 3 and that of the hot wall to 6. The system is composed
of 500 granules. The (red) plus signs correspond to the momentum conserving system, the (blue) stars
to the momentum dissipating system.}
	\label{fig:kappaVisc}
\end{figure}

\begin{figure}[ht]
	\begin{center}
		\includegraphics[width=6cm, angle=-90]{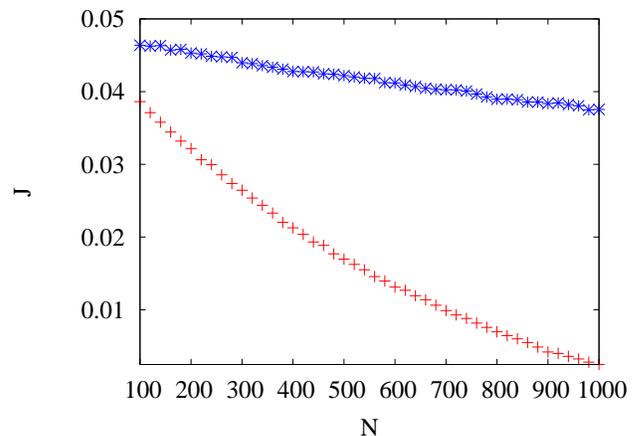}
	\end{center}
	\caption{(Color online) Rates of energy transmission as a function of the system size. In this
simulation the viscosity coefficient was set to $\gamma=0.0001$ and the temperatures of the walls were set to 3 and 6.
The (red) plus signs correspond to the momentum conserving system, the (blue) stars to the momentum
dissipating system.}
	\label{fig:size}
\end{figure}

\begin{figure}[ht]
	\begin{center}
		\includegraphics[width=7cm]{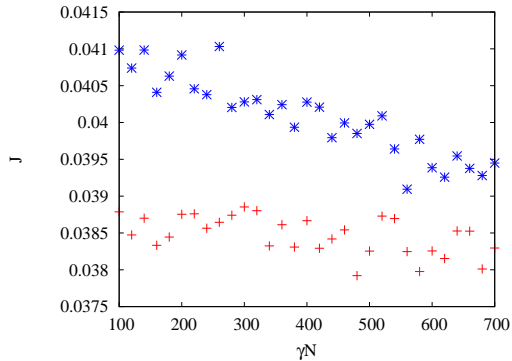}
	\end{center}
	\caption{(Color online) Rates of energy transmission as a function of scaled system size. 
The temperatures of the walls were set to 3 and 6.
The (red) plus signs correspond to the momentum conserving system, the (blue) stars to the momentum
dissipating system.}
	\label{fig:scaled}
\end{figure}

Next, we discuss the dependence of the energy flux on the temperature gradient. The energy flux $J =
d E/d t$
was calculated as the slope of the transmitted energy as a function of time, that is
[(energy injection by the hot wall - energy absorption by the hot wall) - (energy injection by the cold wall -
energy absorption by the cold wall) - (energy dissipated during flow along the chain)] per unit time. As seen
in Fig.~\ref{fig:fourier}, both models lead to behavior fairly well described by Fourier's law.
Repeating this plot for different temperatures of the cold wall, we obtain the dependence of the thermal conductivity
on the temperature. As observed in Fig.~\ref{fig:kappaT}, $\kappa$ is an increasing function of $T$ for both models,
but it is larger for the momentum dissipating system. On the other hand, for a given temperature,
the thermal conductivity decreases (almost linearly) with the viscosity (see Fig.~\ref{fig:kappaVisc}).
Once again, the dissipating system presents larger values of $\kappa$.

Finally, we have analyzed the dependence of the rate of energy transmission on system size, for a fixed temperature
gradient and fixed viscosity. This is the crucial test of normal vs anomalous behavior.
In Fig.~\ref{fig:size} we can clearly see that as the size of the system increases,
the rate of energy transmission does not increase with system size. In fact, it \emph{decreases}. Therefore,
in our model the thermal conductivity does not diverge with increasing system size
regardless of momentum conservation or dissipation. One may be tempted to think of this behavior
entirely as a consequence of the energy dissipation because for bigger systems the number of collisions necessary for
the energy to be transmitted from one end of the system to the other increases. Hence more energy is dissipated
and less energy arrives at the cold wall.  However, the introduction of dissipation changes the
dynamics more profoundly.  This can be seen in Fig.~\ref{fig:scaled}, where we plot the rates of
energy transmission against $\gamma N$, thus taking into account the ``simple" effect of energy
dissipation [see discussion preceding Eq.(\ref{delimit})]. The rate of energy transfer in
the momentum conserving system is essentially
independent of system size in this scaled representation, but that of the momentum dissipating system
actually decreases even when scaled in this way. In any case, we note the very small scale of variation of
the ordinate in Fig.~\ref{fig:scaled}.

\section{Conclusions}
\label{conclusions}

Heat transport in one-dimensional discrete systems continues to be a problem of theoretical
interest, uncertainty, and even controversy~\cite{narayan02,lepri03,dhar08,lepri03b,mai06,mai07}.
Very recently, experimental results in this arena have also started to appear.  In~\cite{geminard05} the focus
is the understanding of transport of heat along colliding granular beads in a liquid medium where
the questions of interest involve the nature of the contact regions (``liquid bridges") between granules.
In~\cite{chang08} the issue is the breakdown of Fourier's law in nanotube thermal conductors in that
the thermal conductivity diverges with length of the nanotube.  The point is that even after many
years of study the conditions that lead to the validity or violation of Fourier's law are not yet
clear.  While momentum conservation has often been featured as a condition closely associated with
the system size divergence of the $1d$ thermal conductivity, we have examined two dissipative
granular gases, one of which involves momentum dissipation along with energy dissipation,
whereas in the other momentum is conserved.  In both of these the thermal conductivity remains
finite and in fact decreases as the number of granules in the system increases, 
as it must in a system where each collision leads to energy dissipation. 
In the momentum dissipating model, it decreases more rapidly than can be accounted for by the simply
heat loss mechanism of the collisions, indicating a more profound change in the dynamics.
This behavior points to the caution 
that must be exercised when associating momentum conservation with anomalous behavior in one
dimension.

\section*{Acknowledgments}

A.R. and I.L.D.P. acknowledge support from Pronex-CNPq-FAPESQ and CNPq.
Acknowledgment is made to the Donors of the American Chemical Society Petroleum Research Fund for
partial support of this research (K.L.).

\end{document}